\documentclass[12pt]{article}

\title{The Physical Principles of Quantum Mechanics. A critical review}

\author{F. Strocchi \\ INFN, Sezione di Pisa, Italy}
\date{}

\makeglossary

\newtheorem{Theorem}{Theorem}[section]
\newtheorem{Definition}[Theorem]{Definition}
\newtheorem{Proposition}[Theorem]{Proposition}

\usepackage{latexsym}
\usepackage[T1]{fontenc}
\usepackage{amsmath}
\usepackage{amssymb}

\def \lan   {\langle}
\def \ran   {\rangle}

\def \AO {{\cal A}({\cal O})}
\def \AO' {{\cal A}({\cal O}')}

\def \Pf {{\bf Proof.\,\,}}

\def \be {\begin{equation}}
\def \ee {\end{equation}}
\def \ume {{\scriptstyle{\frac{1}{2}}}}
\def \ra {\rightarrow}

\def \eqq {\equiv}

\def \a {{\alpha}}
\def \b {{\beta}}

\def \d {{\delta}}

\def \l {{\lambda}}

\def \om {{\omega}}

\def \A {{\cal A}}
\def \B {{\cal B}}

\def \F {{\cal F}}

\def \H {\mbox{${\cal H}$}}
\def \J {{\cal J}}

\def \M {{\cal M}}

\def \O {{\cal O}}
\def \P {{\cal P}}

\def \id {{\bf 1 }}

\def \d^nu {{\partial^\nu}}
\def \d^la {{\partial^\lambda}}
\def \d^o {{\partial^0}}

\def \Cbf {{\bf C}}

\def \Nbf {{\bf N}}
\def \Rbf {{\bf R}}

\def\doppio#1{{\rm I}\kern-.1667em{\rm #1}}

\def\Q{\text{Q}\kern-.52em
    \text{\vrule height1.5ex width.5pt depth0pt}\kern.45em}

\def\dZ{{\mathchoice {\hbox{$\Ss\textstyle Z\kern-0.4em Z$}}
{\hbox{$\Ss\textstyle Z\kern-0.4em Z$}} {\hbox{$\Ss\scriptstyle
Z\kern-0.25em Z$}} {\hbox{$\Ss\scriptscriptstyle Z\kern-0.2em
Z$}}}}

\def\dC{{\mathchoice{\hbox{$\rm\textstyle\text{\kern.35em\vrule
   height1.5ex width.5pt depth0pt\kern-.35em C}$}}
{\hbox{$\rm\textstyle\text{\kern.35em\vrule
   height1.5ex width.5pt depth0pt\kern-.35em C}$}}
{\hbox{$\rm\scriptstyle\text{\kern.35em\vrule
   height1.5ex width.3pt depth0pt\kern-.35em C}$}}
{\hbox{$\rm\scriptscriptstyle\text{\kern.35em\vrule
   height1.5ex width.2pt depth0pt\kern-.35em C}$}}}}

\makeatletter

\@addtoreset{equation}{section}

\begin{document}

\maketitle

\begin{abstract}
The standard presentation of the principles of quantum mechanics is critically reviewed both from the experimental/opera\-ti\-o\-nal point   and with respect to the request of mathematical consistency and logical economy. A simpler and more physically motivated formulation is discussed. The existence of non commuting observables, which characterizes quantum mechanics with respect to classical mechanics, is related to operationally testable complementarity relations, rather than to uncertainty relations. The drawbacks of Dirac argument for canonical quantization are avoided by a more geometrical approach.
\end{abstract}  


\section{The Dirac-von Neumann axioms}

The seminal papers by Heisenberg (1925) and by Schr\"odinger (1926)  mark the birth of quantum mechanics (QM) with apparently different points of view.  Heisenberg setting emphasizes the non-commu\-ta\-tive operator-matrix structure, whereas Schr\"odinger wave mechanics relies on analogies with optical phenomena. The need of unifying such philosophically different approaches is at the basis of Dirac formulation (1930)  of the principles of QM, whose mathematical consistency and refinement is due to von Neumann (1932). The resulting principles  became known as the {\em Dirac-von Neumann (DvN) axioms of quantum mechanics}. We shall briefly review them, pointing out the lack of {\em a priori} physical motivations, as acknowledged by Dirac himself (``The justification of the whole scheme depends on the agreement of the final results with experiments``).  One of the reasons of such a  discussion is that in most  textbook presentations of the principles of  QM, no attempt is made of improving the presentation of the axioms of QM from the experimental/operational point of view (see Section 2).

  The basic idea is Dirac realization of the linear structure of the quantum states, the so-called {\em superposition principle}. It codifies the distinctive feature  of Schr\"odinger wave mechanics with respect to  standard classical mechanics (CM). As stressed by Dirac, in classical wave phenomena  the coefficient $c_\a$ of the $\a$ component in the   superposition describes the weight  of the $\alpha $ contribution  to the final average result of  the measured  observables. On the other hand, in the quantum superposition the coefficient $c_\a$  of the $\a$ state  gives the probability {\em amplitude} for the outcome corresponding to the $\a$ state, i.e. the probability $p_\a$ for the $\a$ outcome is given by $|c_\a|^2$. 
Dirac principles of quantum mechanics (Dirac 1930) may be summarized in the following form, which takes into account  von Neumann revisitation.

\vspace{1mm} \noindent {\bf AXIOM I. States.} {\em The states $\omega  $ of a quantum mechanical system are described by the  rays  $ \underline{\Psi_\om} \eqq \{ \l \Psi_\om, |\l| = 1\}$, identified by  the normalized state vectors $\Psi_\om$,  of a separable Hilbert space
 $\H$.  
More  generally,  a state is described by a density matrix}  $\rho_\om = \sum_j \l^\om_j P_j,\,\,\,\,\l^\om_j \geq 0, \,\,\,\,\,\sum_j \l^\om_j = 1,$ ({\em not all} $\lambda_j$ {\em different}), {\em with $P_j$, $j = 1, ...\infty$, one-dimensional  orthogonal projections in $\H$}. 

As discussed by Dirac, such an axiom  formalizes the {superposition principle} by realizing the underlying structure of vector space spanned by the states. 
The physical basis of such a principle is taken from the analysis of photon polarization experiments, which however do not provide a clear cut distinction with respect to the classical wave picture. The whole Chapter I of Dirac's book is devoted to explaining and possibly justifying such an axiom, which however, in Dirac's words, ``cannot be explained in terms of familiar physical concepts''. 

\vspace{1mm} \noindent {\bf AXIOM II. Observables.} {\em The observables of a quantum mechanical system, i.e. the quantities which can be measured, are described by the set of bounded self-adjoint operators in a Hilbert space  $\H$.}

In Dirac formulation, the operators describing  observables were not required to be bounded and no distinction was made between hermiticity and self-adjointness. 
However,  for an unbounded operator hermiticity is not enough for defining its spectrum and continuous functions of it; hence, its physical interpretation is not well defined. Moreover, the sum of two unbounded self-adjoint operators does not define a self-adjoint operator (see the extensive discussion in Reed and Simon (1975) and therefore  without the condition of  boundedness the whole linear structure of the observables is in question. 
Actually, as shall  argue below, by the intrinsic  limitations of the instruments,   one can measure only bounded functions of self-adjoint operators and for bounded operators hermiticity implies self-adjointness.

A  trivial consequence of axiom II   is that,  through their complex linear combinations and products,  the observables generate an algebra $\A$ over the complex numbers, briefly called the {\em algebra of observables} which coincides with  the whole set $\B(\H)$ of  bounded operators in $\H$. Thus, except for the case of one-dimensional $\H$, the observables generate a {\em non-commutative} algebra. 
The algebra  $\A$ has the following properties:  i) is closed under the involution defined by the adjoint operation, i.e. if $A \in \A$ also $A^* \in \A$, and ii) is equipped with a norm $|| \cdot ||$,  which satisfies  $|| A\,B || \leq || A ||\,\,|| B ||$ and  the so-called $C^*$-{\em property}, 
$||A^* \,A|| = || A ||^2.$ Furthermore, iii) $\A$ is norm closed, i.e. every Cauchy sequence with respect to the norm has a limit which belongs to $\A$. 
An algebra with the above properties i)-iii) is called a  $C^*$-{\em algebra}.

A physically more relevant modification of II turned out to be necessary, in order to account for {\em superselection rules} (Wick, Wightman and Wigner 1952, Wightman 1992), which occur when  there are operators, called {\em superselected charges},  {\em which commute with all the observables}. Notable examples are the electric charge $Q$, the operators which describe the permutations of identical particles, the rotations of angle $2 \pi$ if $\H$ contains states with both integer and half-integer spin etc. In the known cases, the superselected charges have discrete spectra and the Hilbert space $\H$ decomposes as  the direct  sum of subspaces $\H_i$, $i$ running over a discrete index set, called {\em superselection sectors}, each providing an irreducible representation of the algebra of observables. 
The physical implication is that not all projections are observable and that it is impossible  to  measure {\em coherent} superposition of states belonging to different superselection sectors. Hence, axiom II must become


\vspace{1mm} \noindent {\bf AXIOM II. Observables.} {\em The observables are described by the real vector space generated by a subset of the bounded self-adjoint operators in $\H$. }
 
From a technical point of view, the above form of II is equivalent to the statement  that {\em the observables generate a $C^*$-algebra $\A \subseteq \B(\H)$}. 

In Dirac presentation, the physical motivations for the descrip\-tion of the observables by self-adjoint operators look rather weak. Dirac arguments in support of axiom II are somewhat interlaced with implicit assumptions about the spectrum of the observables and its relation with the outcomes of  measurements. 

In the standard presentation of the principles of QM
such an axiom appears as the distinctive feature of QM with respect to classical mechanics (CM) and no {\em a priori} physical motivation is given. Actually, as we shall argue below, axiom II may be justified on the basis of operational considerations and does not characterizes the departure from CM.

 As a consequence of axiom I and II, the states $\om$ define {\em positive linear functionals} on the algebra of observables by $ \mbox{Tr}\,(\rho_\om A), \,\forall A \in \A$. The following axiom relates such  functionals to the experimental expectations $\lan A \ran_\omega$ 
 defined by the limit
of the average of the outcomes of repeated measurements (with the same experimental apparatus) $m^\om_i(A)$, $i = 1, ...$ on {\em identically prepared states},  $ \lan A \ran_
\om \eqq \lim_{N \ra \infty} N^{-1} \,\sum_{i = 1}^N \,m^\om_i(A).$ 
 
The (fast) convergence of the average is needed for the very foundation of experimental physics; for Dirac it is  ``a law of nature'' and for von Neumann ``follows from the so-called law of large numbers, the theorem of Bernoulli''.


\vspace{1mm} \noindent {\bf AXIOM III. Expectations.} {\em If the state $\omega$ is represented by the vector $\Psi_\om \in \H$, then,   for any observable $A$,  the experimental expectation $ \lan A \ran_\omega$,   is given by the Hilbert space matrix element} 
$ \lan A \ran_\omega =  (\Psi_\om, \,A\,\Psi_\om)$. {\em More generally, if $\om$ is represented by the density matrix} $\rho_\om$,  $ \lan A \ran_\omega =  \mbox{Tr}\,(\rho_\om \,A).$   
 
The assertion that the experimental expectations have a Hilbert space realization may look a very strong assumption with no classical counterpart. As  a matter of fact, we shall argue below that  the whole set of axioms I-III applies to the mathematical description of both  QM and CM, the distinctive feature  of QM being solely the non-Abelianess of the algebra of observables, which is encoded in the following axioms IV, V.

\vspace{1mm} \noindent {\bf AXIOM IV. Dirac canonical quantization.} {\em The operators which describe the  canonical coordinates $q_i$ and momenta $p_i$, $i = 1,...s$, of a quantum system with $2 s $ degrees of freedom, obey the  canonical commutation relations} \be{[\,q_i, \,q_j\,] = 0 =[\,p_i,\, p_j\,],\,\,\,\,\,\,\,\,\,\,\,\,\,\,[\,q_i, \,p_j\,] =  i \hbar \,\,\, \delta_{i\,j}.}\ee 

This axiom reflects the commutation relations of the infinite matrices for the position and momentum proposed  by Heisenberg (1925) and later related  to the uncertainty principle (Heisenberg 1927, 1930).
The attempts to justify such an axiom by Heisenberg and by Dirac shall be critically discussed in Sects.\,2.3,2.4.
Eqs.\,(1.1) imply that the canonical variables cannot be described by bounded operators and therefore are not observables according to Axiom II; however, bounded functions of them are candidates for describing observables (see Sect.\,2.5) and their  commutation relations are induced by eqs.\,(1.1).

The following  axiom provides the bridge between Heisenberg and Schr\"{o}\-dinger formulations of QM, a deep open problem at the birth of QM.
The compatibility of the two  descriptions has been the subject of philosophical debates; the recognition that a quantum particle has multiple properties which look contradictory and mutually exclusive has led Bohr (Bohr 1927, 1934) to the formulation of his {\em complementarity principle} as the basic feature of  quantum physics.  

Bohr's  statement is not  mathematically precise and it is not sharp enough to lead to a unique interpretation: `` evidence obtained under different experimental conditions cannot be comprehended within a single picture, but must be regarded as complementary in the sense that only the totality of the phenomena exhausts the possible information about the objects''. This is probably the origin of the still lasting philosophical debates on its meaning.

The following axiom  provides the mathematical formulation of the coexistence of the particle and the wave picture and, together with axiom IV, may be regarded as the substitute of Bohr principle.

  \vspace{1mm} \noindent {\bf AXIOM V. Schr\"odinger representation.} {\em The commutation relations (1.1) are represented by the following operators in the Hilbert space} $\H = L^2(\Rbf^s, d x)$: 
\be{q_i \,\psi(x) = x_i \psi(x), \,\,\,\,\,\,\,\,p_j \,\psi(x) = - i \hbar\, \frac{\partial \psi}{\partial x_j}\,(x).}\ee

\def \Sch  {{Schr\"{o}\-din\-ger\,}}

The lack  of a clear distinction between the role of the  two sets of axioms, I, II, III and IV,V, is at the origin of the widespread point of view, adopted by many textbooks, by which all of them are characteristic of quantum systems.
 The distinction between classical and quantum systems is rather given by the mathematical structure of  $\A$ and it will have different realizations depending on the particular class of systems.


\section{The physical principles of quantum mechanics}

The Dirac-von Neumann axioms provide a neat mathematical foundation of quantum mechanics, but their {\em a priori} justification is not very compelling, their main support, as stressed by Dirac, being  the {\em a posteriori} success of the theory they lead to. 
The dramatic departure from the general philosophy and ideas  of classical physics may explain  the many attempts of obtaining quantum mechanics by a deformation of classical mechanics or by the so-called geometric quantization. Thus, a more argued motivation on the basis of physical considerations is desirable (Strocchi 2005, 2008) and this is the purpose of this section.

\subsection{States, observables and measurements}  

The discussion of the principles of QM gets greatly simplified, from a conceptual point of view, if one first clarifies what are the objects of the mathematical formulation.

The preliminary basic consideration is that Physics deals with reproducible phenomena analyzed by experiments and that any experiment  presupposes a protocol for defining identical  conditions. In fact, an experiment consists of: i) the preparation of the system under consideration in what may be called the initial  state and ii) the measurement of some property or observable of the system  in the so-prepared state. 
Hence,   a physical system is characterized by a set of (initial) states, defined by the corresponding preparation procedures  (Busch,Grabowski and Lahti 1995); with the improvement of the experimental technology a larger number of initial states may possibly be prepared.  

Without such a protocol of {\em preparation of states}, it is impossible  to adopt the statistical interpretation of measurements,  at the basis of experimental physics. This is needed for the definition of  identical conditions and of von Neumann ensembles consisting of replicas of the system in identically prepared states (von-Neumann 1932, pp.\,297, 308).

A physical quantity or {\em observable} $A$ of a physical system is identified by the actual experimental apparatus which is used for its measurement (e.g. the electric current measured by a given amperometer, the magnetic field measured by a given magnetometer etc.). 

The above characterization of preparable states and observables refers to specific  processes of preparation  and to specific instruments;  in order to establish a language exportable to different observers or to different realizations of instruments, {\em equivalence relations} must be acknowledged. 
Thus, if two states defined by two apparently different preparation procedures  yield the same results of measurements for {\em all } observables, i.e. the same expectations, from an experimental point of view they cannot be considered as physically different, since there is no measurement which distinguish them.
Similarly, if two observables $A, B$  defined by two apparently different instruments have the same expectations on {\em all} the preparable states of the system, there is no available operational way of distinguishing them. 

The above considerations may be summarized in the following (phy\-sical) definition of the objects of the physical science  in terms of their operational definitions (quite generally, independently of QM).

\begin{Definition} {\bf States and observables.}   A {\bf physical system} is operationally characterized by the collection $\Sigma$ of all the {\bf states} $\om$, in which it can be prepared according to well specified processes of preparation.

An {\bf observable} $A$ of a physical system is defined by the actual experimental apparatus used for its measurements; a result of  measurement of $A$ in a  state $\om$ is given by the average of the outcomes of repeated measurements $m^\om_i(A)$, $i = 1, ...$ on identically prepared states. The  experimental {\bf expectation} $\lan A \ran_\om$ of $A$ in the state $\om$ is defined by  \be{\lan A \ran_\om \eqq \lim_{N \ra \infty} N^{-1} \,\sum_{i = 1}^N \,m^\om_i(A) ,}\ee the existence of the limit  being  part of  the foundation of experimental phys\-ics. 
The set of observables is denoted by $\O$.

The following equivalence relation is induced by the completeness of the physical operations: \be{ \lan A \ran_{\om_1} = \lan A \ran_{\om_2}, \,\,\,\forall A \in \O, \,\,\, \Rightarrow \,\,\om_1 = \om_2,}\ee  
An observable $A$ is said to be {\bf positive} if all the outcomes of measurements of $A$, $m^\om_i(A)$, are positive for any state $\om$; by the definition (2.1), this implies that  $\lan A \ran_\om \geq 0$, $\forall \om$, i.e. the states define positive functionals on the observables. 
 \end{Definition}

The outcomes $m_i^\omega(A)$ are understood to be obtained by using the experimental apparatus which identifies $A$, and one could spell this out by the more pedantic notation $(m_A)_i^\omega(A)$, which we avoid for simplicity. 

By the above definition,  if  
an instrument $I_A$ identifies the observable $A$,   the (realizable) instrument $I_P$, whose pointer scale is a rescaling of that of $I_A$, in such a way  that the pointer readings of $I_P$ are a given  polynomial function $P$ of the pointer readings of $I_A$,  defines an observable which may be denoted by $P(A)$. This means that $m^\om_i(P(A)) \eqq P(m^\om_i(A))$, $\forall \om$. 

Clearly, the observable $A^2$ is defined by $m^\om_i(A^2) = (m^\om_i(A))^2 \geq 0$, $\forall 
\om$ and therefore it is a positive observable. The observable $A^0$ is  defined  by $m^\om_i(A^0) = (m^\om_i(A))^0 = 1$, (it is the variable defined by property {\bf A} in von Neumann book 1932, p.\,308), and it has the meaning of the identity $\id$, in the sense that 
for any polynomial $P$ one has $\id P(A) = P(A) \id = P(A)$. 
For two polynomials $P_1(A)$, $P_2(A)$, the  product $P_1(A) \,P_2(A)$ is defined by $m^\om_i(P_1(A)\, P_2(A)) = m^\om_i(P_1(A)) m^\om_i(P_2(A))$, $\forall \om$.

  The above operationally motivated definition of the objects of the theoretical description leads  to a rather tight mathematical structure, i.e. each observable generates a $C^*$-algebra (Props.\,2.2, 2.3).

\begin{Proposition} Each observable $A \in \O$ defines a (commutative) polynomial algebra over the reals, $\A_A \subset \O$,   with identity ${\bf 1}$,  equipped with a norm $||\, \cdot \,||$, which satisfies, $\forall  B, C  \in \A_A, \,\,\forall \l \in \Rbf$,  \be{ || B + C || \leq || B ||  + || C ||, \,\,\,\,\,|| \l B || = |\l|\,|| B ||,}\ee \be { || B^2 || \leq ||B^2 + C^2 ||,\,\,\,\,\,\,\,|| B \, C  || \leq || B^2 ||^\ume\, || C^2 ||^\ume.}\ee 
The states define linear (positive) functionals $\om(A)$ on each $\A_A$, continuous with respect to the norm, $| \om(A) | \leq || A ||$. 
\end{Proposition}
\Pf\, By definition of $\A_A$,    $ \forall B,\, C \in \A_A$, $\forall 
\lambda  \in \Rbf$, one has:  

\noindent {\bf m1.} $m^\om_i(\l  B + C  ) = \l \,m^\om_i (B) + m^\om_i (C)$; 
{\bf m2.}  $m^\om_i ( B C ) = m^\om_i (B) \,m^\om_i (C)$;
{\bf m3.}  $m^\om_i (B^2) \geq 0$.

\noindent By definition of expectations  $\lan \lambda B  + C \ran_\om = \lambda \lan B \ran_\om + \lan C \ran_\om$, $\lan B^2 \ran_\om \geq 0,$ i.e. the states define {\em positive linear functionals} $\om(A) \eqq \lan A \ran_\om$ on $\A_A$. Since each observable $A$ is defined by an apparatus with inevitable limitations on the possible outcomes (e.g. the set of positions which the pointer may  take is bounded), independently of the state in which the measurement is done,  one has $ \mbox{sup}_\om |m^\om_i(A)| \eqq m(A) < \infty$.  Then,   \be{ || A || \eqq \mbox{sup}_\om \,|\om(A)| }\ee is a finite positive  number. Obviously, $|\om(A)| \leq || A ||$. 
Furthermore, m1 and the definition (2.5) easily give eqs.\,(2.3) and, by the positivity of the states, the first of eqs.\,(2.4).  
Finally, by the positivity of the states,   $ 0 \leq \om((B + \l C)^2)$, $\forall \l \in \Rbf$, one has $$|\om(B \,C)| \leq  \om(B^2)^{1/2}\, \om(C^2)^{1/2} \leq || B^2 ||^{\ume}\, || C^2 ||^{\ume}, $$ which implies the second of eqs.\,(2.4).   
QED

\vspace{2mm}
It is part of the experimental wisdom that the final results of measurements of observables are recorded by the expectations $\lan A \ran_\omega$ and therefore in order to distinguish two observables $A, B$ one should be able to prepare states $\omega$ such that $\lan A \ran_\omega \neq \lan B \ran_\omega$. Technically,  this means that {\bf the states separate the observables} (just as the observables separate the states), i.e. $$ 
 \lan A \ran_\om = \lan B \ran_\om, \,\,\,\forall \om \in \Sigma, \,\,\, \Rightarrow A = B.$$ 
A weaker form  of such condition, adopted in the sequel, is \be{ 
 \lan A^n \ran_\om = \lan B^n \ran_\om, \,\,\,\forall n, \,\forall \om \in \Sigma, \,\,\, \Rightarrow A = B.}\ee 

\def \AC {\A_A^C}
\vspace{1mm} Further mathematical structure of the observables may be displayed if the states characterize the positivity of the observables.
According to the above definition, the positivity of $A$ requires the positivity of all the outcomes $m^\om_i(A)$, $\forall \om \in \Sigma$, which implies the positivity of  the expectations $\lan A \ran_\om$, $\forall \om \in \Sigma$, 
but does not seem to be implied by it. However, the role of the states in completely identifying the observables suggests that 
 one may prepare enough states so that if some outcome of measurements of $A$ is not positive, one can prepare a state $\omega$ such that $\lan A \ran_\om$ is not positive. This motivates the following assumption:

\vspace{1mm}\noindent {\bf Completeness of states for the   positivity of observables}. \be{ \lan A \ran_\om \geq 0, \,\,\,\,\, \forall \om  \in \Sigma, \,\,\,\,\mbox{implies}\,\,\,\, m^\om_i(A) \geq 0, \,\,\,\,\forall \om,}\end{equation} 
or in the weaker form $$\lan A^n \ran_\om \geq 0, \,\,\,\,\,\forall n, \, \forall \om  \in \Sigma, \,\,\,\,\mbox{implies}\,\,\,\, m^\om_i(A) \geq 0, \,\,\,\,\forall \om. $$
The two conditionsare equivalent if the positive functionals 
$$ \Omega_{A^n}(A) \eqq \lim_{N \ra \infty} N^{-1} \sum_{i = 1}^N m_i^\omega(A^n A A^n)$$ describe the expectations of realizable states $\omega_{A^n}$, since then $\lan A \ran_\omega \geq 0$, $\forall \omega$ implies $\lan A^{2n + 1}\ran_\omega = \lan A^{2n} \ran_\omega \, \lan A \ran_{\omega_{A^n}} \geq 0$. 

\begin{Proposition} Under the above assumption (2.7)
each observable $A$ generates a (commutative)  $C^*$-al\-ge\-bra $\A_A^C$, with identity,  through the  complex polynomials of $A$,  and the set theoretical union of such algebras $\A_A^C$ yields an extension $\O^C$ of $\O$; the states are naturally extended  to linear positive functionals on each $\A_A^C$.
\end{Proposition}
\Pf\, We start by showing that  
$|| B^2 || = || B ||^2$, $\forall B \in \A_A$. The second of eqs.(2.5) gives
 $|| B^2 || \geq || B ||^2$. 
On the other hand, by eq.\,(2.6),  $\om(|| B || \id \pm B) \geq 0$, $\forall \om$, so that, by eq.\,(2.7),    $|| B || \id \pm B $ are positive elements of $\A_A$. 
Then,  by property m2, one has  $$m^\om_i(|| B ||^2 \id - B^2)  = m^\om_i(|| B || \id + B)\,m^\om_i(|| B || \id - B) \geq 0;$$ hence,  $\om(|| B ||^2 \id - B^2) \geq 0 $, which implies  $ || B ||^2 \geq || B^2 || $. Hence,  
\be{|| B^2 || = || B ||^2.}\ee

\noindent Furthermore, by eq.\,(2.8), the second of eqs.\,(2.5) becomes 
\be{|| B \, C  || \leq || B ||\, || C ||,\,\,\,\,\,\,\forall B \in \A_A.}\ee

\noindent As a next step,  for any pair  $B, \,C \in \A_A$, one may introduce the complex combination $B + i \,C$, operationally defined by taking the corresponding complex combination of the outputs in the measurements of $B$ and $C$. 
In this way one gets a complex extension $\AC$ of $\A_A$, which is stable under the involution $^*$ defined by $(B + i \,C)^* = B - i \,C$, $ \forall B, \,C \in \A_A$.  
\noindent Clearly, m1, m2 still hold for the elements of $\A_A^C$ and furthermore, by definition one has, $\forall B \in \AC$, 

\noindent {\bf m4.} $m^\om_i(B^*) = \overline{m^\om_i(B)}$,\,\,
{\bf m5.} $m^\om_i(B^*\, B) \geq 0.$    

\noindent The   extension of the norm from $\A_A$ to $\AC$ is defined in such a way that the $C^*$-condition is satisfied  
 \be{ || B || = || B^* \,B ||^{1/2}.  }\ee 

\noindent The analog of eq.\,(2.9) is obtained by noting that, by the definition of the norm in $\AC $ (and commutativity),  for $G, \,H \in \AC$, one has $$ || G\, H ||^2  =  || H^* G^*\,G H || = || H^* H \, G^* G || \leq $$  $$\leq || H^* H||\, || G^* \,G|| = || H ||^2 \,|| G ||^2, $$ where eqs.\,(2.10), (2.9)  have  been used for  $H^* H, G^* G \in \A_A$.

\noindent For the triangle inequality, one first notes that, by eqs.\,(2.8), (2.5), (2.6), for $ G \eqq  B + i \,C$, $H \eqq D + i \,E$, $B, C, D, E, \in \A_A$, one has  $$ \ume || G^* H + H^* G ||^2 =  || B D + C E ||^2  = || (B D + C E)^2 || \leq $$ $$ \leq || (B D + C E)^2 + (B D - C E)^2 || \leq  || (B^2 + C^2) (D^2 + E^2) || \leq $$ \be{ \leq  || B^2 + C^2 || \, || D^2 + E^2 || =   || G ||^2 \,|| H ||^2.}\ee  
$$|| G + H ||^2 = || (G + H)^* \,(G + H) || = || G^* G  +  G^* H + H^* G + H^* H || \leq $$
$$ \leq || G^* G || + || G^* H + H^* G || + || H^* H || \leq || G ||^2  + 2 || G ||\, || H || + || H ||^2 =  $$ 
$$ = ( || G || + || H ||)^2, $$
where  eq.\,(2.6) has been used for $G^* G,   G^* H + H^* G,   H^* H$, all belonging to $ \A_A$, and the above derived inequality (2.11). 


\noindent  From a technical point of view and without loss of generality it is  convenient to consider the norm completion of $\AC$, still denoted by $\AC$, to which the states have a unique extension by continuity. In this way one gets a commutative    $C^*$-algebra. The set theoretical union of such norm closed commutative algebras is still denoted by $\O$.
 QED

 \vspace{2mm} For several pairs of observables $A, B \in \O$, (not necessarily commuting, like e.g. $S_1^2$ and $S_2^2$,  where $S_i, i = 1, 2, 3$ are the components of a spin one and $S_1^2 + S_2^2$ is the observable $S^2 - S_3^2$),  there exists an observable $C \in \O$ such that \be{\om(C) = \om(A) + \om(B), \,\,\,\,\forall \om \in \Sigma,}\ee so that, by  eq.\,(2.6) and the commutative addition of the  expectations, one may introduce the commutative addition $ A + B = C$.
  
In general, given a pair $A, B$ one may introduce a (commutative)  addition by eq.\,(2.12); however, the so introduced $C$ is not guaranteed to share all the properties which the observables have as a consequence of their operational definition. One might want to consider the set $\O$  equipped only with a partial addition operation; however, it seems natural to assume that $\O$ may be embedded in a larger set, where the sum of any two observables has the properties of an observable. This qualifies as an axiom beyond the implications of the operational analysis discussed so far; however, in our opinion, represents a more physically motivated alternative to Dirac-von Neumann axiom II. An indirect justification of it  as a  property of the description of a general physical system is that it is satisfied by {\em  both CM and QM}. All the preceding discussion and arguments are meant to provide  a physical justification of such an axiom and are completely summarized and superseded by it. 

\vspace{2mm} \noindent  {\bf AXIOM A. Algebra of observables. } {\em The  observables generate a $C^*$-algebra $\A$, with identity,  briefly called {\bf algebra of observables};  the {\bf states}, which, by eq.\,(2.1),  define positive linear functionals on the algebras $\A_A \subset \A$, for any observable $A$, separate such algebras in the sense of eq.\,(2.6) and extend to positive linear functionals on $\A$.} 

  

\subsection{The Hilbert space realization of states and observables}
The next step is the mathematical realization of the states and observables. 
The Dirac-von Neumann axioms I and III follow from axiom A, through the 
Gelfand-Naimark-Segal (GNS) theorem (Gelfand and Naimark 1943, Segal 1947); thus, 
 quite generally {\em the experimentally defined states, 
observables and expectations have a Hilbert space description}.

\def \ho {\H_\om}
\def \po {\pi_\om}
\def \pso {\Psi_\om} 

\begin{Theorem}
 A state $\om$, identified by its  expectations 
on the $C^*$-al\-ge\-bra of observables  $\A$, defines a Hilbert space $\H_\om$, 
 a representation $\pi_\om $ of the observables as bounded operators in $\H_\om$   and a cyclic vector $\Psi_\om \in \H_\om$, such that the expectations are represented by the Hilbert space matrix elements $ \om(A) = (\pso, \,\po(A)\,\pso), \,\,\,\,\forall A \in \A.$

\noindent The triplet $(\ho, \po, \pso)$ is uniquely determined up to isomorphisms.
\end{Theorem}

The proof recognizes that a state $\omega$ defines a strictly positive inner product on the vector space $[\A] = \A/\J$, being the ideal of null vectors, and $\H_\om$ is obtained by completion of $[\A]$. The representation $\pi_\om$ is explicitly defined by $\pi_\om([A]) \,[B] = [AB]$, $\forall [A], [B] \in [\A]$. Putting $\Psi_\om = [\id]$ one gets $ \om(A) = (\pso, \,\po(A)\,\pso)$.    

The representations $\po$  of $\A$ by algebras of operators in a Hilbert space are not necessarily {\em faithful}, i.e. there may be elements  $A \in \A$ which are mapped to zero, $\po(A)=0$; the following   Theorem (Gelfand and Naimark 1943) states that the algebra of observables can be identified with (i.e. it is faithfully represented by)   a $C^*$-algebra of operators in a Hilbert space $\H = \oplus_{\om \in \F}\,\ho$,  the    
direct sum of the Hilbert spaces $\ho$ defined by a {\em separating family} $\F$ of states, i.e. such that for any $0 \neq A \in \A$ there is at least one $\om \in \F$, with $\om(A) \neq 0$ .
\begin{Theorem}(Gelfand, Naimark) The $C^*$-algebra of observables is isomorphic to a $C^*$-algebra of  operators in a Hilbert space $\H$.
\end{Theorem} 

\Pf A faithful representation $\pi$ of $\A$ is obtained by considering the direct sum $\H$ of the Hilbert spaces $\ho$ defined by a family $\F$ of states, which separate the observables in the sense of eq.\,(2.6), i.e. such that for any non zero element $A \in \A$ there is at least one $\om \in \F$ and an $n \in \Nbf$, with $\om(A^n) \neq 0$: $\H = \oplus_{\om \in \F}\,\ho$. The vectors $x \in \H$ have components $x_\om \in \ho$, such that $|| x ||^2 \eqq \sum_{\om \in \F} ||x_\om||^2 < \infty.$

\noindent The representation  $\pi$ is defined by $\pi(A) = \oplus_{\om \in \F}\, \po(A)$ and $\pi(A) \in \B(\H)$, since  $$ ||\pi(A)\,x||^2 = \sum_{\om \in \F} || \po(A) x_\om||^2 \leq || A ||^2  \sum_{\om \in \F} || x_\om ||^2. $$ Furthermore, for any non zero $A$, $\pi(A) \neq 0$, since, if $A \neq 0$ , by eq.\,(2.6) for at least one $\om$ and an integer $n$, $\pi_\omega(A^n)= \pi_\omega(A)^n \neq 0$, which implies $\pi_\omega(A) \neq 0$.  QED

\subsubsection{Superposition principle and non-Abelian algebra of observables}


\vspace{1mm} Dirac discussion and motivation of axioms I, II, make  crucial reference to the superposition principle and the Gelfand-Naimark theorem, which states that the algebra of observables is isomorphic to a $C^*$-algebra of operators,   may look surprising, being derived from axiom A, independently of QM.
The point is that the  superposition of states leads to quantum effects only if the algebra of observables is non-Abelian and the distinctive property of the classical systems  is the Abelianess of $\A$. 
\begin{Proposition}The relative phase in the superposition of two states of the Gelfand-Naimark  Hilbert space $\H$ (belonging to different rays) is observable only if the algebra of observables is non-Abelian.
\end{Proposition}
\Pf \, The Hilbert space of an  irreducible representation  $\pi$ of an Abelian $C^*$-algebra $\A$ is one-dimensional, since irreducibility requires that the operators which commute with $\pi(\A)$ are multiples of the identity and therefore  $\pi(A) = \l(A) \id $, with $\l(A) \in \Cbf$, by Abelianess of $\A$.  
Such representations are defined  by multiplicative  states $\omega$ (i.e. such that $\om(A\,B)= \om(A)\, \om(B)$), since by the GNS theorem  $\om(AB) = (\id,\,\pi_\omega(A) [B]) = \l_\omega(A) \omega(B) = \omega(A) \omega(B)$. Conversely, multiplicative states define one-dimensional representations.
Clearly, the family $\F$ of all multiplicative states is separating for $\A$ and, by  
a general result (Gelfand and Naimark 1943), it has the structure of a compact Hausdorff space, with the weak * topology $\tau$. Then, the GN Hilbert space  is $\H  = \oplus_{\omega \in \F} \H_\omega$, and $\pi(A) =\oplus \pi_\omega(A)$, $\pi_\omega(A) = \lambda_\omega(A)\, \id$, is equivalently identified by the collection $\{ \lambda_\om(A) = \tilde{A}(\omega), \,\omega \in \F \}$, i.e. by the $\tau$ continuous function $\tilde{A} \in C(\F)$. Thus, the GN representation is equivalently given by  the Abelian algebra  $C(\F)$. 
 
\noindent For example, in the case of classical Hamiltonian systems, with bounded phase space $\Gamma$, $\A$ is isomorphic to the $C^*$-algebra of continuous functions  $C(\Gamma)$ on $\Gamma$, with the sup norm.   
 The  expectations of the multiplicative states are the averages defined by $\delta$ functions, supported by the points $P \in \Gamma$: $\om_P(A) = \tilde{A}(P)$, $\tilde{A} \in C(\Gamma)$. 

\noindent The important physical consequence is that in the GN Hilbert space a non-trivial superposition defines a mixed state on $\A$, 
 the relative phases  being  not observable since the irreducible representations of $\A$ are one-dimensional.
\goodbreak

\subsection{Complementarity  and non-commuting obser\-va\-bles}
\vspace{2mm}

An important issue discussed by  the founders of QM, especially by Bohr and Heisenberg, was 
the identification of  a basic principle which characterizes QM with respect to CM. 
This problem, with foundational and philosophical implications, led Bohr to the formulation of his {\em complementarity principle} (Bohr 1927, 1934): ``adopt a new mode of  description designated as complementary in the sense that any application of classical concepts precludes the simultaneous use of other classical concepts  which in a different connection are equally necessary for the elucidation of the phenomena". The somewhat ambiguous wording of such a principle, also in subsequent presentations and discussions by Bohr, caused considerable disagreement in the literature  about its actual meaning, especially  in regard of its role of providing the conceptual foundations  of QM. The main problem is the precise interpretation of the principle in terms of unambiguous experimental operations and its precise mathematical formulation. 

A more definite and concrete step was taken by  Heisenberg (Heisenberg 1930) with the proposal of basing the principles of QM on the {\em uncertainty relations}, or {\em  uncertainty principle}, i.e. on the experimental limitations which inevitably affect the measurement of position and momentum of a quantum particle, more generally the measurements of canonically conjugated variables
 \be{\Delta_\om(q)\,\Delta_\om(p) \geq \hbar/2, \,\,\,\,\,\,\,\,\,\forall \om,}\ee 
where $\Delta_\om(A)^2 \eqq \om(\tilde{A}^2)$, $ \tilde{A} \eqq A - \om(A)$.

 Bohr and  most of the following literature considered the Heisenberg uncertainty relations  as the ``symbolic expression'' of the complementarity principle, since two canonically conjugated  variables provide an example of complementary descriptions in terms of each of such variables,  which cannot take sharp values on the same state.

 Since it is the existence of non-commutating observables which characterizes QM with respect to CM, according to the general philosophy adopted so far, one should trace this property back to experimental facts. Contrary to a widespread belief, it is only at this stage that QM enters in the Dirac-von Neumann axioms.

The standard argument for the non-Abelianess of the algebra generated by the canonical variables $q, p$ makes reference to Heisenberg uncertainty relations. 
The argument  
makes use of the Robertson inequality (Robertson 1929) for any two pairs of hermitian operators (neglecting domain problems) \be{\Delta_\om(A)\,\Delta_\om(B) \geq \,\ume |\om([\,A,\,B\,])|,}\ee implied by  the positivity of the states, $\om((\tilde{A} - i \l\,\tilde{B})\,( \tilde{A} + i \l\,\tilde{B}) \geq 0$. 

Clearly, if one takes $A = q$, $B = p$, the Heisenberg uncertainty relations follow if $q, p$ obey the canonical commutation relations. 
On the other hand, on the basis of Robertson inequalities, the canonical commutation relations qualify as the natural algebraic structure responsible  for the Heisenberg uncertainty relations (Heisenberg 1927, p.\,174-175). This is the physical motivation given by Heisenberg in his book (Heisenberg 1930).

 Heisenberg discussed {\em gedanken} experiments, with the aim of  supporting or even proving the uncertainty  relations, but it seems that he was aware of the difficulty of turning the {\em gedanken} experiments into real experiments. Clearly,  an unquestionable {\em indirect support} of the canonical commutation relations  is provided by the success of QM which relies on them, but for the foundations of QM one would like to have a direct experimental proof of the Heisenberg uncertainty relations.

Actually, recent attempts to test the quantum lower limit of the product $\Delta_\omega(q)\,\Delta_\omega(p)$ have faced non-trivial difficulties (Nairz, Arndt and Zeilinger 2002). As a matter of fact, one may wonder whether  uncertainty relations of the form \be {\Delta_\om(A)\,\Delta_\om(B) \geq \,C > 0,\,\,\,\,\,\,\forall \om, }\ee  are   compatible with $A$ and $B$ being bounded operators, as required  by their interpretation as observables. 

To this purpose, we remark that, 
given an observable $A$, the inevitable limitations in the preparation of states  and in the measurements of $A$  in general preclude the possibility of obtaining sharp values for $A$, i.e. $\Delta_\omega(A) = 0$. However, according to the present wisdom  of experimental physics, the following principle  is tacitly taken for granted: 
\vspace{2mm}

\noindent {\bf Experimental principle. Preparation of states and  sharp values of observables.} {\em For any  given observable $A$ one can correspondingly  prepare states for which a sharp value may be approximated as well as one likes, i.e.} \be{ \mbox{Inf}_\om \,\Delta_\om(A) = 0.}\ee 

If $\A$ has more than one physical (irreducible) representation, as for quantum systems with an infinite  degrees of freedom, the above lower limit should hold for the states in each of such representations. 

Such an experimental principle implies that the {\em uncertainty relations of the form (2.15) cannot hold for a pair of (operationally defined) observables $A, \,B$}.
In fact, since the operational definition of observables implies that they are described by bounded operators, one has $$ \Delta_\omega(B)^2 = \omega(B^2) - \omega(B)^2 
\leq   2 \,|| B ||^2,$$ and $$ \mbox{Inf}_\omega \,\Delta_\om(A)\,\Delta_\om(B) \leq \mbox{Inf}_\omega\Delta_\om(A)\,\sqrt{2}\, || B || = 0.$$
This  means that  it is impossible to have a direct experimental check of the uncertainty relations (2.13), since one only measures bounded functions of the position and of the momentum.
  
The impossibility of preparing states $\omega$ for which the measurements of two observables $A, B$ give (almost) sharp values may be better formalized by replacing the notion of uncertainty by the following notion, which we propose to call {\bf complementarity}.

\begin{Definition} Two observables $A, \,B$ are called {\bf complementary} if the following bound holds $$  
 \Delta_\om(A) + \Delta_\om(B) \geq \,C > 0,\,\,\,\,\forall \,\om .$$ 
\end{Definition}

This provides a  precise {\em operational and mathematical formulation of complementarity} with the advantage, w.r.t.  the Heisenberg uncertainty relations, of being meaningful and therefore testable for operationally defined observables, necessarily represented by bounded operators. 
As we shall argue below, under general assumptions the existence of complementary observables, in the sense of Definition 2.7, implies that the  algebra of observables cannot be commutative, i.e. the distinctive  property of quantum systems with respect to classical systems.

 The relation between complementarity and non-com\-mu\-ta\-ti\-vity  is easily displayed if one realizes that  in each  irreducible representation $\pi(\A)$ of the algebra of observables  one may enlarge the notion of observables by considering  as observables the weak limits of any Abelian $C^*$-subalgebra $\B \subset \pi(\A)$. 
Technically, this amounts to  consider the von Neumann algebra $\overline{\B}^w$ generated by $\B$; one may show that $\overline{\B}^w$ contains all the spectral projections of the elements of $\B$. 
 In the Gelfand representation of the Abelian $C^*$-algebra $\B$ by the set of continuous functions on the spectrum of $\B$, such weak limits correspond to the pointwise limits of the continuous functions. They are operationally defined by instruments whose outcomes yield the pointwise limits of the functions defined by  the measurements  of the  elements of $\B$. 

This means that one recognizes as observables not only the polynomial functions of elements $ B \in \B$, and therefore by norm closure the continuous functions of $B$, but also their pointwise limits. 
 \begin{Proposition} If the above experimental principle holds, given a representation $\pi$ of $\A$, the existence of  two  observables $\pi(A), \pi(B)$ which are complementary, implies that  the $C^*$-algebra $\A(A,B)$ generated by $\pi(A), \pi(B)$  cannot be commutative
\end{Proposition}
\Pf Clearly, $\A(A,B)$ is commutative if and only if so is the  von Neumann algebra $\overline{\A(A,B)}^w$ and in this case, by   a theorem by von Neumann (von Neumann, Ann. Math.  1931) there exists $C \in \overline{\A(A,B)}^w$ such that $\pi(A)$ and $\pi(B)$ may be written as functions of $C$. Then, the states which  yield (almost) sharp values   for $C$ (whose existence  follows from the above experimental principle, eq.\,(2.16)) also yield (almost) sharp values for $\pi(A)$ and $\pi(B)$ and complementarity cannot hold. QED

A way out of the implication of non-commutativity from complementarity is to take the somewhat {\em ad hoc } point of view that  there are observables for which (2.16) does not hold. This is implicitly at the basis of the attempts of  formulating QM in terms of a commuting algebra of observables. 

The non-commutativity of the algebra of observables for quantum system can be traced back to the existence of a pair of complementary observables, i.e. to the check of the quantum bounds in the inequality of complementarity. 
The following examples provide simple  experimental tests of complementarity relations, supported by experimental evidence.\goodbreak

\vspace{2mm}
\noindent {\bf Example 1. Spin 1/2.} 
 
The spin components, say $s_1, s_3$  are not conjugate variables and their commutator may have vanishing expectation; therefore Robertson inequality for the derivation of uncertainty relations is not useful. In fact, $\inf_\om \Delta_\om(s_1) \Delta_\om(s_3) = 0$. This is an example in which the complementarity relations  prove to be  more effective than uncertainty relations. 
In fact, the most general two-component spin state $\chi = (\chi_1, \chi_2)$ may be reduced to either one of the following forms $$\chi_1 = 1/N, \,\chi_2 = (x + i y)/N, \,\,\,\,\,\,\,\,\,\chi_1 = (x + i y)/N, \, \chi_2 = 1/N,$$ with $ N^2 = 1 + x^2 + y^2$, and the commutation relations $[\,s_1,\, s_3 \,] = - i\,\hbar s_2$  imply that  on both states the  sum of the above mean square deviations   is $(\hbar^2/4)( 1 + 4 y^2 )$.   
Hence, one has the following  quantum bound  
$$  (\Delta_\om(s_1))^2 + (\Delta_\om(s_3))^2 = \ume  \hbar^2 - \omega(s_x)^2 - \om(s_3)^2 \geq \,\hbar^2/4 > 0.$$  
Since all the experimental measurements of the expectations of the spin variables are in agreement with the quantum theoretical predictions, the above quantum bound  may be regarded an experimentally established fact, and $s_1, s_3$ are complementary observables.

\vspace{1mm}\noindent {\bf Example 2. Weyl operators.} 

Since one can measure only bounded functions of $q, p$ one cannot directly test eq.\,(2.13). 
However, we shall argue that there are  bounded functions of $q,  p$ for which the quantum limit predicts  complementarity, and  one may test the quantum bounds. 

To this purpose, we consider the Weyl exponentials $U(\a) = \exp{i \a q}$, $V(\b) = \exp{i \b p}$, which are elements of the  algebra generated by the observables of a quantum particle. Then, putting  $ \tilde{q} \eqq \sqrt{s/\hbar}(q - \om(q)), \,\,\tilde{p} \eqq (\hbar \,s)^{-1/2}(p - \om(p)),$ with $s$ a suitable parameter with the dimensions of a mass times an inverse time,  we look for a lower bound of $ (\Delta_\om(\cos \tilde{q}))^2 + (\Delta_\omega(\cos \tilde{p}))^2.$ 

\def \tilq {\tilde{q}}
\def \tilp  {\tilde{p}}

Since we are looking for states with expectations of $q, p$ as sharp as possible,  the expectations of $\tilde{q}^n$, $\tilde{p}^n$ must be as small as possible. Thus,  we may estimate  the above mean square deviations for  $\tilde{q}$, $\tilde{p}$ small  $$ (\Delta_\om(\cos \tilde{q}))^2 + (\Delta_\omega(\cos \tilde{p}))^2 \sim  {{\scriptstyle{\frac{1}{4}}}} [( \om((\tilq^2) - \om(\tilq^2))^2) + \om((\tilp^2 - \om(\tilp^2))^2)].$$
On the other hand, by positivity, for any hermitian operator $A$ one has $$\om(A^2) - \om(A)^2 = \om((A - \om(A))^2) \geq 0,\,\,\,\,\,\mbox{ i.e.}\,\,\,\, \om(A^2) \geq \om(A)^2.$$ Hence, one has $$ \om((\tilq^2) - \om(\tilq^2))^2) + \om((\tilp^2 - \om(\tilp^2))^2) \geq \Delta_\om(\tilq)^4 + \Delta_\om(\tilp)^4.$$ The quantum limit set by  the Heisenberg uncertainty relations gives $\Delta_\om(\tilp) \geq  \ume (\Delta(\tilq))^{-1}$ and one gets the following  quantum lower bound $$(\Delta_\om(\cos \tilde{q}))^2 + (\Delta_\omega(\cos \tilde{p}))^2  \geq $$ $$  1/4 [( \om((\tilq^2) - \om(\tilq^2))^2) + \om((\tilp^2 - \om(\tilp^2))^2)] \geq (\ume)^3.$$ An experimental  test of such a bound will prove that the algebra generated by the Weyl exponentials cannot be commutative.

\subsection{Dirac canonical quantization} 
The extraordinary success of the emerging theory raised  the question of   
a derivation  of the canonical commutations relations from general principles; a related problem is the relation between classical mechanics and quantum mechanics, since the mere limit $\hbar \ra 0$ recovers commutativity, but the canonical structure is lost. As a response to such questions, 
in his 1930 book, Dirac discussed  a derivation of the canonical commutation relations by the  method, or rather the principle  of {\em classical analogy} (Dirac 1930), leading to what became known as {\bf Dirac canonical quantization}.  

As a first step, Dirac noticed that the (regular) functions $f(q, p)$ of the canonical variables (including the constants) generate an associative  algebra $\A_{cl}$, with the associative product $(f g)(q, p) \eqq f(q, p) g(q, p)$,    which is also equipped with the  product $ \{ \cdot, \, \cdot \}$ given  by the (classical) Poisson brackets $ \{ f, \, g \} $.
Such a product satisfies the properties of a Lie product: i) linearity in both factors, ii) antisymmetry, iii) the {\em Leibniz rule with respect to the associative product} $$\{ f,\, g\,h \} = \{ f, \,g\} h + g \{ f,\, h \},$$ and iv) the Jacoby identity $ \{ f, \,\{ g,\, h \} \} + \{ h, \,\{ f,\, g \} \} + \{ g, \,\{ h,\, f \} \} = 0.$
The  principle of {\em classical analogy} assumes that also  the quantum variables define an associative algebra $\A_q$, equipped with a Lie product $\{ \cdot, \, \cdot \}$ satisfying i)- iv) and that there is a correspondence between the classical and the quantum variables $ f \ra u_f$ such that a)  the correspondence is linear, 
b) the   constant functions  $c \in \Cbf$ are mapped into multiples of the identity $c \id$,  and 
c) the following relation holds between the Lie products in $\A_{cl}$ and in $\A_q$ \be{ \{ \,u_f, \,u_g \} = u_{\{ f,\, g \}}.}\ee
The important idea underlying Dirac strategy is the realization that both the classical and quantum algebras are Poisson algebras, i.e. have an associative product (which defines the commutator $A B - B A$) as well as a Lie product which  satisfies the Leibniz rule with respect to the associative product and gives  the canonical structure. 

As a second step, Dirac proved that in a Poisson algebra $\A$ the associative and the Lie products  are related by the {\bf Dirac identity} \be{ (A B - B A) \,\{ C, \, D\} = \{ A,\, B\}\, (C D - D B), \,\,\, \forall A, B, C, D, \in \A.}\ee 
Such an identity, rediscovered by Farkas and Letzer  in the general context of Poisson algebras (Farkas and Letzer 1998), follows merely from linearity and the Leibniz rule satisfied by the Lie product with respect to the associative product (by comparing the results of expanding $\{ A C, \,B D \}$ with the Leibniz rule, first applied $A C$ and then to  $B D$ and secondly in reverse order).

As a final step, Dirac argued that,  since eq.\,(2.18)  holds  with $A$ and $B$ quite independent of $C$ and $D$, we  must have \be{( A B - B A ) = i \,\hbar\, \{ A, \,B \}, \,\,\,\,(C D - D C) = i \,\hbar \, \{ C, \, D \}, }\ee
with $\hbar$ independent from $A, B,  C, D$ and commuting with $A B - B A$. 

Irreducibility then implies that $\hbar$ is a c-number and one has a strong relation between the commutator defined by the associative product and the Lie product which gives the canonical structure. 

We have dissected Dirac argument in order to discuss its weak and good points. The first weak point  is the assumed correspondence $\A_{cl} 
\ra \A_q$ satisfying a)-c). 
In fact, in an irreducible representation of the quantum algebra linearity implies that the {\em von Neumann rule}  holds for polynomial functions $\P(q, p)$, i.e. $u_{\P(q,p)} = \P(u_q, u_p)$, and this leads to inconsistencies (Groenwold 1946). 
This problem is at the origin of a rich  literature about possible improvements or alternatives to Dirac quantization, with results which in most cases end up displaying  mathematical obstructions or no-go theorems (see the review by Ali and Englis 2005).
Moreover, Dirac identity for Poisson algebras requires  additional assumptions  for the derivation of  the  crucial eq.\,(2.19). 

As remarked by Farkas and Letzer there are Poisson algebras for which such a relation does not hold,  a sufficient condition being that the Poisson algebra is prime, i.e. it does not have ideals which are divisors of zero; however, such a condition is not satisfied by the algebras of functions of canonical variables on a manifold and Dirac argument is not complete.
A possible way of overcoming Dirac weak points is to give up the idea of relating  the classical and the quantum algebras through a mapping satisfying  a)-c) and rather obtain them as different realizations of a unique  Poisson algebra. 
To this purpose,  one may consider the free polynomial (associative) algebra $\Lambda$ generated by the associative (commutative) algebra $C^\infty(\Rbf^s)$ of the  
infinitely differentiable functions $f(q)$ of the coordinates  and by the  momenta $p_j$. 
 The crucial point is that the (canonical)  Lie product $\{ \cdot, \,\cdot \}$  \be{ \{ f(q),\, p_i \} = \partial f /\partial q_i, \,\,\,\,\,\{ f(q),\, g(q) \} = 0 = \{ p_i, \,p_j \}, }\ee is extended to  $\Lambda$  {\em exclusively} through linearity, antisymmetry and the Leibniz rule.  
Thus, $\Lambda$ is  a Poisson algebra and clearly is common to CM and QM, the distinction between the two amounting to the different relation between the commutator and the Lie product. 
There is a unique involution or adjoint operation $^*$ on $\Lambda$, which leaves the canonical variables $q_i$, $p_i$ invariant and satisfies $\{ A^*, \,B^* \} = \{ A, \, B \}^*$. 

It is worthwhile to remark that $\Lambda$ is not the Poisson algebra defined by Dirac, not only because it is not affected by the ordering problem, but especially  because eq.\,(2.17) does not hold and, as we shall see,  its   Lie product reduces to that of the Dirac Poisson algebra only after  a quotient (with respect to a central variable). 
 
 \begin{Theorem} The Poisson algebra $\Lambda$ contains a central variable $Z $, such that, $\forall A, B \in \Lambda$ 
\be{ [ A, \,B \,] \eqq (A B - B A) = Z \{ A, \,B \}, \,\,\,\,\,\{ Z, \,A \} = 0 = [ Z, \, A],}\ee
in particular 
\be{ [ q_i, \,p_j ] = Z  \{ q_i, \, p_j \} = Z \delta_{i \,j},\,\,\,\,\,\,Z = - Z^*, }\ee $$ [ q_i, \,q_j ] = Z \{ q_i,\, q_j \} = 0,\,\,\,\,\,\,[ p_i, \,p_j ]  =  Z \{ p_i,\, p_j \}  = 0.$$
 \end{Theorem}    
The proof exploits not only the Dirac identity (2.19), but also crucially the very special Lie product recursively defined on $\Lambda$ by the canonical Lie products $\{ q_i, \,p_j \} = \delta_{i j }, $ etc. 
In fact, eq.\,(2.19) with $A = q_1$, $B = p_1$, $C = q_i$, $D = p_i$, $i = 1, ...s$ gives
$[ q_1, \,p_1 ] = [ q_i,\, p_i ] \eqq Z. $ 
Moreover, the recursive definition of the Lie product on $\Lambda$  through the Leibniz rule and the canonical Lie products gives $$ \{ q_i, \, Z \} = \{ q_i, \,q_j p_j - p_j q_j \} = 0, \, \,\,\,\,\{ p_i, \, Z \} = \{ p_i, \, q_j p_j - p_j q_j \} = 0,$$ 
so that $\{ Z, \,A \} = 0$, $ \forall A \in \Lambda$. 
Furthermore, the Dirac identity yields $$ Z \{ C,\, D \} = [ q_1, \, p_1 ]\, \{ C,\, D \} = \{ q_1, \, p_1 \} \, [ C,\, D ] = [ C,\, D ], $$   $$\{ C,\, D \} Z =  [ C,\, D ] \{ q_1, \,p_1 \} = [ C, \, D ],$$ i.e. $Z$ commutes with all Lie products and therefore with $ q_i = \ume \{ q_i^2, p_i \}$,  $p_i = \ume \{ q_i, p_i^2 \}$, $i = 1,...s$. Thus,  $Z$ is a central variable with respect to both the Lie product and the commutator $[ A,\, B ] \eqq (A B - B A)$. 

\vspace{1mm} The above Theorem replaces the incomplete Dirac argument for eq.\,(2.19) and shows that classical and quantum mechanics are the only two factorial (i.e. with trivial center) realizations of the {\em same} Poisson algebra $\Lambda$, corresponding to $Z = 0$ and to $Z = i \hbar \neq 0$, respectively.
 
\def \cm {$C^\infty(\M)$}
\def \vm {Vect\,$(\M)$\,}

In view of the above result,  it appears somewhat improper to look for a mapping between the  classical and quantum canonical algebras, just as it would be for two inequivalent representation of a Lie algebra (or group),  and  this partly explains the obstructions for approaches to quantization based on  such a mapping. 
The non-vanishing anti-adjoint  variable $Z$ also explains the {\em need of complex numbers for} the description of {\em quantum mechanics}, in contrast with  the classical case. 
The above approach to canonical quantization can be generalized to the case of an arbitrary configuration manifold $\M$, yielding   a {\em diffeomorphism covariant quantization} (Morchio and Strocchi 2008, 2009). 


\subsection{Schroedinger wave mechanics}

\vspace{1mm} \noindent The last Dirac-von Neumann axiom may be derived under very general mathematical conditions from the canonical commutations relations (1.1). 
To this purpose, it is convenient to work with bounded functions of the canonical coordinates in order to avoid the delicate domain questions as well as possible mathematical pathologies.
The standard choice has become that advocated by Weyl (Weyl 1927, 1950), namely to choose as basic mathematical objects the exponentials of the Heisenberg canonical variables $ q \ra  U(\a) = \exp{( i 
\a q)}$,  $p \ra V(\b) = \exp{( i \,\b p)}$, $\a, \b \,\in 
\Rbf^s $. 
As $C^*$-algebra one takes the {\em unique} (Slawny 1971, Manuceau et al. 1973) $C^*$-algebra $\A_W$, called the {\bf Weyl algebra}, generated by the Weyl operators $U(\a), V(\b)$, $\a, \b \in \Rbf^s$, with the algebraic relations induced by the Heisenberg canonical commutations relations   and by the self-adjointness of $q, 
\, p$: $ U(\a)\,U(\a') = U(\a + \a')$, $V(\b)\,V(\b') = V(\b + \b')$, $U(\a)\, V(\b) = V(\b) \,U(\a) e^{- i \a \b}$, $U(\a)^* = U(-\a)$, $V(\b)^* = V(-\b)$.
The classification of the representations of the canonical commutation relations is then reduced to that of the Weyl algebra $\A_W$.

As it is standard in group theory, the classification of the corresponding representations $\pi$ is done under the  general condition that the representatives $\pi(U(\a)\,V(\b))$ are strongly (equivalently weakly) continuous in the variables $\a, \b$. This is a necessary and sufficient condition for the existence of the generators $q, p$ (Stone 1932). A representation with such a property is called {\em regular}.
  
\begin{Theorem} (Stone 1930, von Neumann 1931) All irreducible regular representations of the Weyl algebra $\A_W$ are unitarily equivalent.
\end{Theorem}
Since the Schr\"{o}dinger representation $\pi_S$ in the Hilbert space $\H = L^2(\Rbf^s, d^s x)$: $(\pi_S(U(\a))\psi)(x) = e^{i \a\,x}\,\psi(x), \,\,(\pi_S(V(\b))\psi)(x) = \psi(x + \b)$ is irreducible and regular, modulo isomorphisms, it is the {\bf unique} irreducible regular representation of the canonical commutation,  relations; thus,  under very general regularity conditions, one  gets a proof of axiom 5,  by  using Axiom A and the algebraic structure induced by the canonical commutation relations.

In conclusion, the operational definition of states and observables motivates the physical principle or axiom that, quite generally  the observables of a physical (not necessarily quantum mechanical) system generate a $C^*$-algebra. The Hilbert space realization of states and observables (Dirac-von Neumann AxiomsI-III) is then a mathematical result. The existence of observables which satisfy the operationally testable complementarity relations implies that the algebra of observables is not commutative and it marks the difference between classical mechanics and quantum mechanics. Thus, for a quantum mechanical system the Poisson algebra generated by the canonical variables cannot be represented by commuting operators and actually canonical quantization  (Axiom IV) follows from general geometrical structures. The Schroedinger representation (Axiom V) is uniquely selected by irreducibility and regularity. 
The general setting discussed so far may then provide a more economical and physically motivated alternative to the Dirac-von Neumann axioms for the foundation of quantum mechanics.

\newpage
REFERENCES
\vspace{2mm}

S.T. Ali and M. Englis, Quantization methods: a guide for physicists and analysts, Rev. Math. Phys. {\bf 17}, 391-490 (2005)

M. Born and P. Jordan, Zur Quantenmechanik,   Zeit. f. Physik, {\bf 34}, 858-888 (1925); English translation in  {\em Sources of Quantum Mechanics}, B.L. van der Waerden ed., Dover 1968, paper 13

N. Bohr, The quantum postulate and the recent development of atomic energy, talk at the Como Conference 1927, published in Nature(Supplement), {\bf 121}, 580-590 (1928); reprinted in {\em Atomic Theory and the Description of Nature}, Cambridge University Press 1934 

P. Busch. M. Grabowski and P.J. Lahti, {\em Operational Quantum Physics}, Springer 1995

P.A.M. Dirac, {\em The principles of Quantum Mechanics}, Oxford University Press 1930

D.R. Farkas and G. Letzer, Ring theory from symplectic geometry, Jour. Pure Appl. Algebra {\bf 125}, 155-190 (1998)

G.B. Folland, {\em Harmonic Analysis in Phase Space}, Princeton University Press 1989

I. Gelfand and M.A. Naimark, On the imbedding of normed rings into the ring of operators in a Hilbert space, Mat. Sborn., N.S., {\bf 12}, 197-217 (1943)

H.J. Groenwold, On the principles of elementary quantum mechanics, Physica {\bf 12}, 405-460 (1946)

H. Hanche-Olsen and E. St{\o}rmer, {\em Jordan Operator Algebras}, Pitman 1984

W. Heisenberg, Uber quantentheoretische Umdeutung kinematischer und mechanischer Beziehungen, Zeit. f. Physik,  {\bf 83}, 879-893 (1925); English translation in {\em Sources of Quantum Mechanics}, B.L. van der Waerden ed., Dover 1968, paper 12

W. Heisenberg, Uber den anschaulichen Inhalt der quantentheoretischen Kinematik und Mechanik, Zeit. f. Physik {\bf 43}, 172-198 (1927)

W. Heisenberg, {\em The Physical Principles of Quantum Theory}, University of Chicago Press 1930

P. Jordan,   \"{U}ber die Multiplikation quantenmechanischen G\"{o}ssen, Zeit. f. Physik, {\bf 80}, 285-291 (1933)

P. Jordan, \"{U}ber Verallgeinerungsm\"{o}glichkeiten des Formalismus der Quantenmechanik, Nach. Acad. Wiss. G\"{o}ttingen, Math. Phys. Kl. 1, {\bf 41}. 209-217 (1933)

P. Jordan, J. v. Neumann and E. Wigner, On an algebraic generalization of quantum mechanical formalism, Ann. Math. {\bf 35}, 29-64 (1934)

J. Manuceau, M. Sirugue, D. Testard and A. Verbeure, The smallest $C^*$-algebra for canonical commutations relations, Comm. Math. Phys. {\bf 32}, 231-243 (1973)

G. Morchio and F. Strocchi, The Lie-Rinehart Universal Posson Algebra of Classical and Quantum Mechanics, Lett. Math. Phys. {\bf 86}, 135-150 (2008)

G. Morchio and F. Strocchi, Classical and quantum mechanics from the universal Poisson-Rinehart algebra of a manifold, Rep. Math. Phys. {\bf 64}, 33-48 (2009)

O. Nairz, M. Arndt and A. Zeilinger,  Experimental verification of the Heisenberg uncertainty principle for fullerene molecules,   Phys. Rev. {\bf A 65},  032109-1-4   (2002)

M. Reed and B. Simon, {\em Methods of Modern Mathematical Physics}, Vol.\,II, (Fourier Analysis, Self-adjointness), Academic Press 1975

H.P. Robertson, The Uncertainty Principle, Phys. Rev. {\bf 34}, 163-164 (1929)

E. Schr\"odinger, Quantisierung as Eigenwertproblem, Ann. Physik, {\bf 79}, 361-376; 489-527; {\bf 80}, 437-490; {\bf 81}, 109-139 (1926)

E. Schr\"odinger, An undulatory theory of the mechanics of atoms and molecules, Phys. Rev. {\bf 28}, 1049-1070 (1926)

I. Segal, Postulates of general quantum mechanics, Annals of Math., {\bf 48}, 930-948 (1947)

J. Slawny, On factor representations and the $C^*$-algebra of canonical commutations relations, Comm. Math. Phys. {\bf 24}, 151-170 (1972)

M.  Stone, Linear transformations in Hilbert space, III. Operational methods and group theory, Proc. Nat. Acad. Sci. U.S.A. {\bf 16}, 172-175 (1930)
 
M. Stone, On one-parameter unitary groups in Hilbert space, Ann. Math. {\bf 33}, 643-648 (1932)

F. Strocchi, {\em An Introduction to the Mathematical Structure of Quantum Mechanics},  World Scientific 2005, 2nd ed. 2008

J. von Neumann, Die Eindeutigkeitder Schr\"{o}dingerschen Operatoren. Math. Ann. {\bf 104}, 570-578 (1931)

J. von Neumann, Uber Functionen von Functionaloperatoren, Ann. Math. {\bf 32}, 191-226 (1931)

J. von Neumann, {\em  Mathematical Foundations of Quantum Mechanics}, Princeton University Press 1932

H. Weyl, Quantenmechanik und Gruppentheorie, Zeit. f. Physik, {\bf 46}, 1-46 (1927)

H. Weyl, {\em The Theory of Groups and Quantum Mechanics}, Dover 1950

G.C. Wick, A.S. Wightman and E.P. Wigner, Phys. Rev. The intrinsic parity of elementary particles, {\bf 88}, 101-105  (1952) 

A.S. Wightman, Some Comments on the quantum theory of measurement, in {\em Probabilistic Methods in Mathematical Physics}, F. Guerra et al. eds., World Scientific 1992

\vspace{2mm}

\end{document}